\documentclass[twocolumn,amsmath,amssymb,superscriptaddress,prl]{revtex4-2}
\usepackage[utf8]{inputenc}
\usepackage{color}
\usepackage{amsmath}
\usepackage{amssymb}
\usepackage{graphicx}
\usepackage{textcomp}
\usepackage{dcolumn}
\usepackage{bm}
\usepackage[right]{eurosym}
\usepackage{float}
\usepackage[english]{babel}
\usepackage{blindtext}
\usepackage{babel}
\usepackage[normalem]{ulem}

\begin{document}

\title{Single-shot electron imaging of dopant-induced nanoplasmas}

\author{C. Medina} 
\affiliation{Physikalisches Institut, Universit{\"a}t Freiburg, 79104 Freiburg, Germany}

\author{D. Schomas} 
\affiliation{Physikalisches Institut, Universit{\"a}t Freiburg, 79104 Freiburg, Germany}

\author{N. Rendler} 
\affiliation{Physikalisches Institut, Universit{\"a}t Freiburg, 79104 Freiburg, Germany}

\author{M. Debatin}
\affiliation{Physikalisches Institut, Universit{\"a}t Freiburg, 79104 Freiburg, Germany}
\altaffiliation[Present address: ] {Institut f{\"u}r Physik, Universit{\"a}t Kassel, 34132 Kassel, Germany} 

\author{D. Uhl} 
\affiliation{Physikalisches Institut, Universit{\"a}t Freiburg, 79104 Freiburg, Germany}

\author{A. Ngai} 
\affiliation{Physikalisches Institut, Universit{\"a}t Freiburg, 79104 Freiburg, Germany}

\author{L. Ben Ltaief} 
\affiliation{Department of Physics and Astronomy, Aarhus University, 8000 Aarhus C, Denmark}

\author{M. Dumergue}
\affiliation{ELI-ALPS, ELI-HU Non-Profit Ltd., Wolfgang Sandner utca 3., Szeged, H-6728, Hungary}

\author{Z. Filus}
\affiliation{ELI-ALPS, ELI-HU Non-Profit Ltd., Wolfgang Sandner utca 3., Szeged, H-6728, Hungary}

\author{B. Farkas}
\affiliation{ELI-ALPS, ELI-HU Non-Profit Ltd., Wolfgang Sandner utca 3., Szeged, H-6728, Hungary}

\author{R. Flender}
\affiliation{ELI-ALPS, ELI-HU Non-Profit Ltd., Wolfgang Sandner utca 3., Szeged, H-6728, Hungary}

\author{L. Haizer}
\affiliation{ELI-ALPS, ELI-HU Non-Profit Ltd., Wolfgang Sandner utca 3., Szeged, H-6728, Hungary}

\author{B. Kiss}
\affiliation{ELI-ALPS, ELI-HU Non-Profit Ltd., Wolfgang Sandner utca 3., Szeged, H-6728, Hungary}

\author{M. Kurucz}
\affiliation{ELI-ALPS, ELI-HU Non-Profit Ltd., Wolfgang Sandner utca 3., Szeged, H-6728, Hungary}

\author{B. Major}
\affiliation{ELI-ALPS, ELI-HU Non-Profit Ltd., Wolfgang Sandner utca 3., Szeged, H-6728, Hungary}

\author{S. Toth}
\affiliation{ELI-ALPS, ELI-HU Non-Profit Ltd., Wolfgang Sandner utca 3., Szeged, H-6728, Hungary}

\author{F. Stienkemeier}
\affiliation{Physikalisches Institut, Universit{\"a}t Freiburg, 79104 Freiburg, Germany}

\author{R. Moshammer}
\affiliation{Max-Plank-Institut f{\"u}r Kernphysik, 69117 Heidelberg, Germany}

\author{T. Pfeifer}
\affiliation{Max-Plank-Institut f{\"u}r Kernphysik, 69117 Heidelberg, Germany}

\author{S. R. Krishnan}
\affiliation{Department of Physics and QuCenDiEM-group, Indian Institute of Technology Madras, Chennai 600036, India}

\author{A. Heidenreich} 
\affiliation{Kimika Fakultatea, Euskal Herriko Unibertsitatea (UPV/EHU) and Donostia International Physics Center (DIPC), P.K. 1072, 20080 Donostia, Spain}
\affiliation{IKERBASQUE, Basque Foundation for Science, 48011 Bilbao, Spain}

\author{M. Mudrich} 
\affiliation{Department of Physics and Astronomy, Aarhus University, 8000 Aarhus C, Denmark}
\affiliation{Indian Institute of Technology Madras, Chennai 600036, India}
\email[E-mail me at: ]{mudrich@phys.au.dk}

\begin{abstract}
We present single-shot electron velocity-map images of nanoplasmas generated from doped helium nanodroplets and neon clusters by intense near-infrared and mid-infrared laser pulses. We report a large variety of signal types, most crucially depending on the cluster size. The common feature is a two-component distribution for each single-cluster event: A bright inner part with nearly circular shape corresponding to electron energies up to a few eV, surrounded by an extended background of more energetic electrons. The total counts and energy of the electrons in the inner part are strongly correlated and follow a simple power-law dependence. Deviations from the circular shape of the inner electrons observed for neon clusters and large helium nanodroplets indicate non-spherical shapes of the neutral clusters. The dependence of the measured electron energies on the extraction voltage of the spectrometer indicates that the evolution of the nanoplasma is significantly affected by the presence of an external electric field. This conjecture is confirmed by molecular dynamics simulations, which reproduce the salient features of the experimental electron spectra.
\end{abstract}

\date{\today}

\maketitle

\section{Introduction}

The generation of nanoplasmas in clusters and nanoparticles by intense femtosecond laser pulses is being widely studied, both for exploring fundamental light-matter interactions under extreme conditions, and in view of potential applications. Nanoplasmas absorb laser light very efficiently and convert it into fast electrons~\citep{shao1996multi}, highly-charged ions~\citep{snyder1996intense}, and energetic radiation~\citep{saalmann_mechanisms_2006,fennel_laser-driven_2010}. Therefore, nano- and microplasmas bear the potential to serve as sources of extreme-ultraviolet (XUV) and x-ray radiation~\citep{masim2016nanoplasma}, as compact accelerators for charged and neutral particles~\citep{fennel_laser-driven_2010,rajeev2013compact}, and for high-harmonic generation~\citep{Kundu:2007,vampa2015linking}.

The ionization dynamics of clusters and nanodroplets induced by intense near-infrared (NIR) pulses is fairly well understood, owing to dedicated experiments and detailed model calculations~\citep{fennel_laser-driven_2010, saalmann_mechanisms_2006,heidenreich_extreme_2007,heidenreich_charging_2017}. Initiated by tunnel ionization~\citep{rose1997ultrafast}, an ionization avalanche is launched within a few optical cycles mainly by electron-impact ionization, thereby releasing a large fraction of electrons from their parent atoms or ions (inner ionization)~\citep{last2004electron}. These quasi-free electrons remain trapped in the space-charge potential of the cluster and mainly determine the optical response of the resulting nanoplasma. Some electrons are directly emitted by the combined action of the laser and the Coulomb field of individual ions (outer ionization). Depending on the pulse duration or the delay between short pulses in a dual-pulse scheme, resonant absorption occurs, thereby further enhancing the yield, charge state and kinetic energy of emitted ions~\cite{Zweiback:1999,krishnan_evolution_2012}. As the nanoplasma heats and charges up, it starts expanding and electrons evaporate out of the collective Coulomb potential of the cluster core. This leads to a thermal distribution of free electrons characterized by an essentially exponential energy dependence~\citep{pohl2004exponential,ovcharenko_novel_2014,schutte2014rare,schutte_correlated_2015,kelbg_auger_2019}. Finally, slow electrons and ions partly recombine in the course of the expansion. Highly excited neutral and ionic species formed during this phase can decay by correlated electronic decay processes, as recently observed~\citep{schutte_correlated_2015, kelbg_auger_2019,oelze_correlated_2017, kelbg_temporal_2020}.

Helium (He) nanodroplets are a special type of rare-gas clusters due to the extremely high ionization energy of He atoms and the unique superfluid nature of He droplets~\citep{mudrich_photoionisaton_2014}. When doping He droplets with impurity atoms or molecules, the latter tend to aggregate inside the droplet or at the surface where they form a cluster. Electrons emitted from this dopant cluster by tunnel ionization then act as seeds to ignite a nanoplasma in the entire He droplet~\citep{mikaberidze_laser-driven_2009,krishnan_dopant-induced_2011}. 
Thus, doping He droplets with various types of atoms or molecules in an adjustable quantity offers an additional control knob to tune the susceptibility of the droplets to intense NIR pulses~\cite{mikaberidze_laser-driven_2009,krishnan_dopant-induced_2011,heidenreich_dopant-induced_2017}. In particular, the addition of a few dopant atoms with low ionization energy such as xenon (Xe) can drastically reduce the intensity threshold for nanoplasma ignition ~\citep{mikaberidze_laser-driven_2009,krishnan_dopant-induced_2011,Heidenreich:2016, heidenreich_dopant-induced_2017}. 
Furthermore, electron spectra of ionized He droplets are relatively simple to interpret due to the simple electronic structure of He and the large spacing between atomic levels~\citep{kelbg_auger_2019}. 

Experimentally, He nanodroplets can be created in a wide range of sizes spanning many orders of magnitude~\citep{toennies_superfluid_2004, gomez_sizes_2011}. Dopant clusters formed by aggregation are mostly located inside the nanodroplets, the only exception being alkali metals~\citep{Heidenreich:2016}. In contrast, neon (Ne) clusters solid nanocrystallites. The location of dopants picked up by Ne clusters inside or at the surface of the clusters is less well defined~\citep{buck_cluster_1996,ltaief_vuv_2018}. 

He nanoplasmas have mainly been studied using ion detection techniques~\citep{doppner_ion_2006, krishnan_dopant-induced_2011,Heidenreich:2016,heidenreich_dopant-induced_2017}. Recently, time-of-flight (TOF) spectroscopy of electrons has revealed additional details about the dynamics of He nanoplasmas~\citep{kelbg_auger_2019, kelbg_comparison_2018, kelbg_temporal_2020}. However, so far only averaged data over many laser shots and nanoplasma hits have been reported. In this paper, we present for the first time single-shot electron images from individual He nanoplasmas driven by intense NIR and mid-infrared (MIR) femtosecond laser pulses. We show typical examples of individual velocity-map images (VMIs) recorded under different conditions, and we systematically analyze large sets of VMIs to identify important trends and dependencies. By simultaneously recording electron VMIs and ion TOF mass-over-charge spectra we find that electron and ion kinetic energies are strongly correlated by a simple power law.

The salient feature of the VMIs is a mostly centro-symmetric intensity distribution with greatly fluctuating size and brightness for each plasma explosion. We show that this feature can be described by a simple heuristic model based on a homogeneously charged spherical electron cloud. A more complete classical molecular dynamics (MD) simulation shows that the external electric field applied in the VMI spectrometer for extracting electrons and ions influences the electron energy distributions. Simulated velocity distributions and electron spectra are compared with the experimental data.

\begin{figure}[th!]
\centering
\includegraphics[width=1\columnwidth]{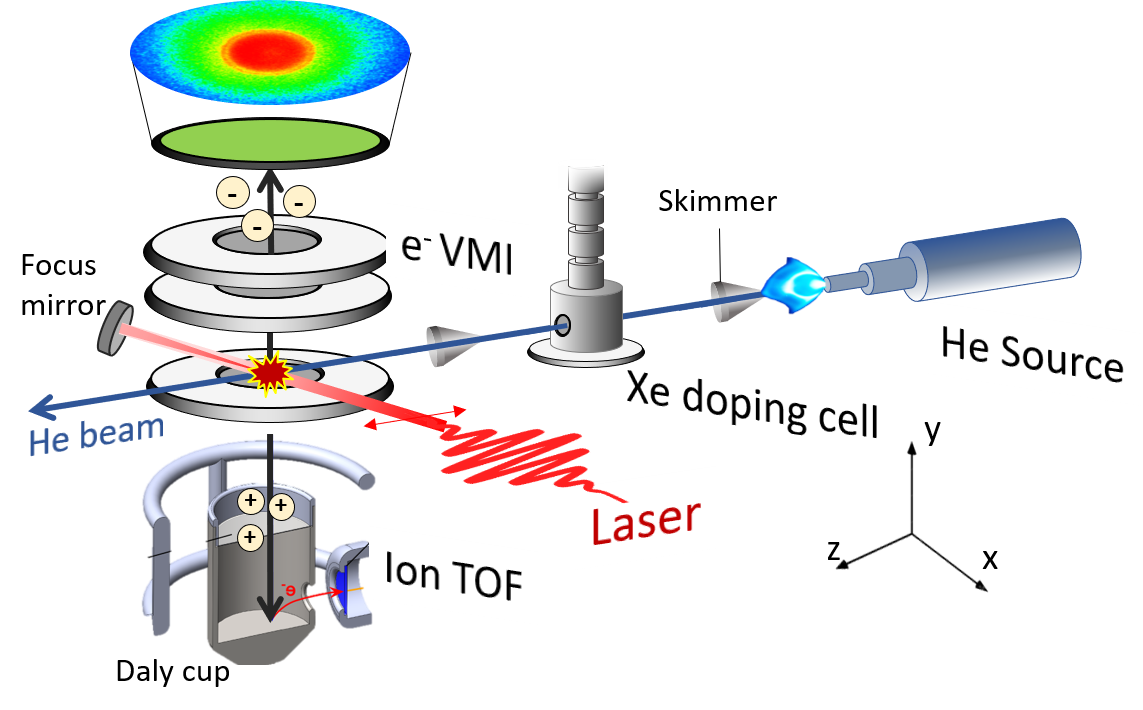} \caption{\label{setup} Schematic illustration of the experimental setup. The He droplet source, doping cell and spectrometer are contained in differentially-pumped vacuum chambers separated by beam skimmers. Electrons are imaged on imagin detector on the top of the spectrometer; ions are detected by a Daly type time-of-flight (TOF) detector on the bottom. The laser beam is back focused into the He droplet beam at right angles by a spherical mirror ($f=75$~mm). }
\end{figure}

\section{Methods}

\subsection{Experimental setup}
The nanodroplet apparatus resembles the one we have used previously~\citep{krishnan_dopant-induced_2011, krishnan_evolution_2012, Heidenreich:2016}. The main novelty is the implementation of a combined VMI and TOF detector, capable of detecting energetic electrons and ions, see Fig.~\ref{setup},~\cite{schomas_compact_2017}. The He nanodroplets, respectively Ne clusters, were produced by a supersonic expansion at high stagnation pressure ($p_{0}=20$-$50$, 10 bar) and low temperature ($T_{0}=9$-$15$~K, $37$-$41$~K) through a thin-walled nozzle with a diameter of $5$~$\mu$m for He and $20~\mu$m for Ne. The size of He nanodroplets was estimated based on literature data~\citep{toennies_superfluid_2004}; the Ne cluster size was calculated using Hagena's scaling law~\citep{hagena_cluster_1992}. Doping with Xe atoms is achieved by passing the clusters through a doping cell containing Xe gas at adjustable pressure, see Fig.~\ref{setup}. The size of the Xe dopant cluster is estimated based on the Poissonian pick-up statistics while taking into account the shrinkage of the He droplets induced by the aggregation of a dopant cluster~\citep{kuma2007laser}.

NIR (800~nm) laser pulses of $35$~fs duration (FWHM) with a repetition rate of 3~kHz are focused into the cluster beam using a back-focusing mirror with a focal length of 75~mm. With a pulse energy of up to 200~$\mu$J, the peak intensity ranges up to $1.5\times10^{15}$~Wcm$^{-2}$ as inferred from the distribution of charge states of strong-field ionized xenon atoms~\citep{augst1991laser}. Most of the measurements presented in this work were measured at a pulse energy of 80~$\mu$J. A few experiments were carried out using a high-repetition-rate (100~kHz) MIR laser centered at a wavelength of $3.2~\mu$m which provides pulses with energy up to 100~$\mu$J and a length of about 6 optical cycles (50~fs)~\citep{thire2018highly}, yielding peak intensities up to $4\times10^{14}$~Wcm$^{-2}$. This laser system was located at the ELI-ALPS Research Institute in Szeged, Hungary. 

The detector consists of a compact VMI spectrometer capable of detecting high-energy electrons~\citep{schomas_compact_2017}, and a TOF ion spectrometer placed on the opposite side which is based on a Daly type detector~\cite{daly1960scintillation}, see Fig.~\ref{setup}. The VMI and TOF spectrometers were synchronized to the laser trigger using a delay generator. One advantage of the single-shot VMI technique is that very short exposure times in the microsecond range can be used, which reduces the background level. In most of the measurements we adjusted the experimental parameters such that on average less than 10\,\% of the images contained a nanoplasma hit to ensure that the rate of images containing two or more hits remained low. Given the tight focusing of the laser beam and the low number density of doped He droplets in the interaction region, multiple hits induced by the same laser pulse can be safely excluded. 

\subsection{Molecular dynamics simulations}
\label{sec:MD}
The molecular dynamics (MD) simulation method for the interaction of a cluster with the electric and magnetic field of a linearly polarized NIR Gaussian laser pulse was described previously \citep{heidenreich_efficiency_2016,heidenreich_extreme_2007,heidenreich_simulations_2007}. Starting from neutral atoms, electrons enter the MD simulation, when the criteria for tunnel ionization (TI), classical barrier suppression ionization (BSI) or electron impact ionization (EII) apply. This is checked at each atom in the course of the trajectory, using the local electric field at the atoms as the sum of the external laser electric field and the contributions from all ions and nanoplasma electrons. Instantaneous TI probabilities are calculated by the Ammosov-Delone-Krainov (ADK) formula~\citep{ammosov_tunnel_nodate}, EII cross sections by the Lotz formula \citep{lotz_empirical_1967}, calculating the ionization energy with respect to the local atomic Coulomb barrier in the nanoplasma environment \citep{fennel_highly_2007}. The electron dynamics is treated relativistically. Three-body electron-ion recombination (TBR) is automatically accounted for in the MD simulations.

\begin{figure*}[th!]
\centering
\includegraphics[width=2.0\columnwidth]{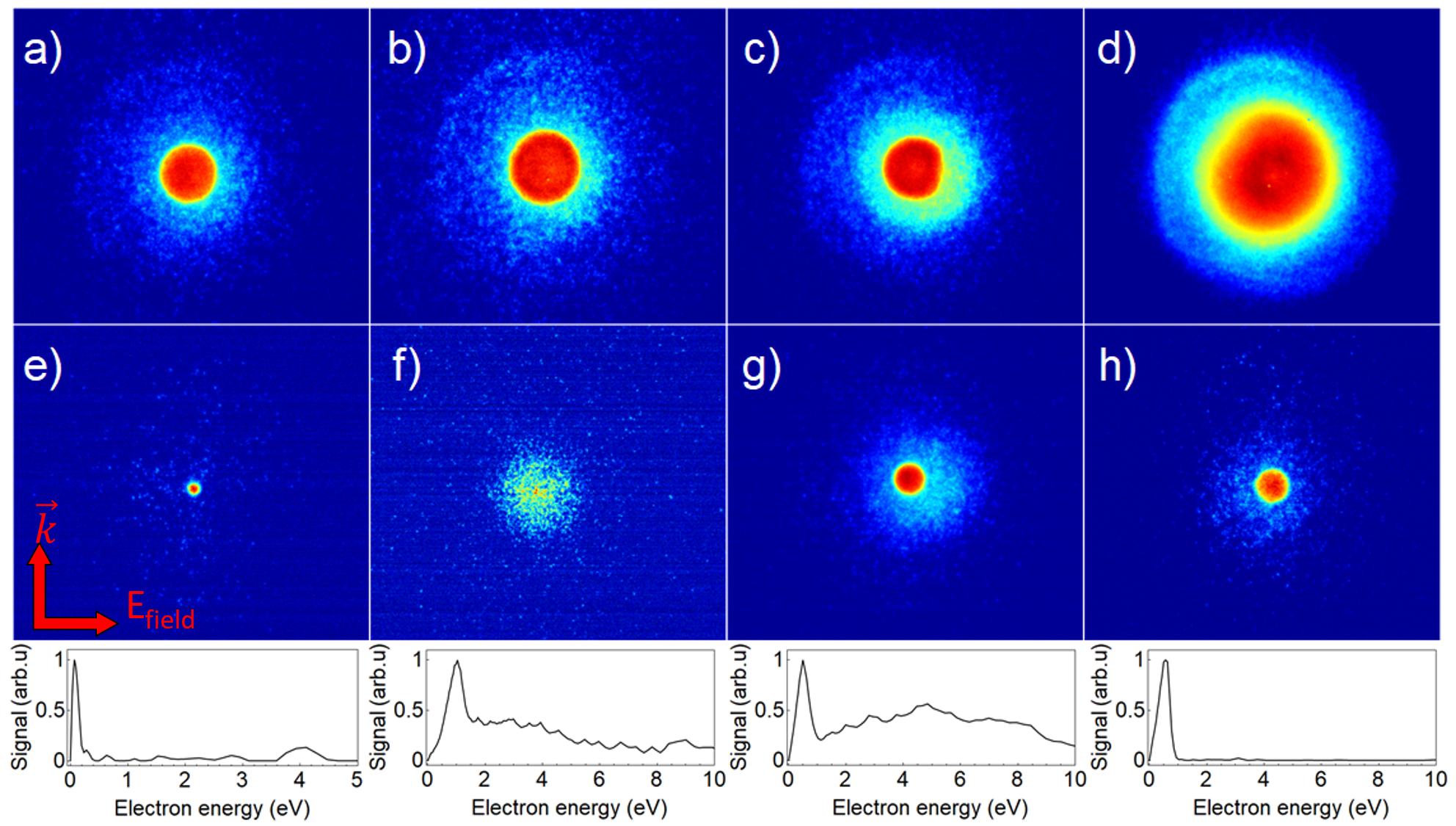} \caption{\label{fig1}Selected single-shot electron VMIs of He nanoplasmas induced by intense NIR laser pulses. a-b) Typical VMIs for large droplets consisting of $2\times10^{5}$ He atoms doped with 40 Xe atoms. e-h) Typical electron VMIs for smaller droplets ($2\times10^3$ He and 8 Xe atoms). The arrows in e) indicate the direction of propagation of the laser ($\vec{k}$) and its polarization parallel to the detector plane ($\vec{E}_{field}$). The He beam propagates parallel to $\vec{E}_{field}$. The bottom row shows electron spectra inferred from the small-droplet images (middle row).}
\end{figure*}

Ion-ion interactions are described by Coulomb potentials, ion-electron and electron-electron interactions by smoothed Coulomb potentials. Interactions between neutral atoms and of neutral atoms with cations are disregarded. Accordingly, the formation of He$_N^+$  cationic complexes as well as collisional kinetic energy transfer between neutral atoms and between neutral atoms and cations is not accounted for. The Pauli repulsive potential between electrons and neutral He atoms is taken as a sum of pairwise forth-order Gaussian functions located at every neutral He atom, $V\left(r_{ij}\right)=V_0\exp\left(-r_{ij}^4⁄\sigma^4\right)$, where $r_{ij}$ is the He-electron distance, $V_0 = 1.1$~eV \citep{Buchenau:1991}, and the exponent $\sigma = 1.2$~\AA~is chosen such that the effective range of $V_{ij}$ is about half of the He-He distance ($3.6$~\AA) of the neutral droplet. The Xe-Xe distances are $4.33$~\AA~(bulk), He-Xe distances $4.15$~\AA~\citep{chen_intermolecular_1973}. The simulations are carried out for He$_{2171}$Xe$_{23}$, NIR pulse peak intensities $I = 2\times10^{14}$~Wcm$^{-2}$, and a FWHM pulse duration of 35~fs. For these laser parameters, the He droplet is nearly completely ionized; the ionization degree depends very little on the initial conditions of the trajectories, so that averaging over a set of trajectories that is usually necessary in case of doped He droplets \citep{heidenreich_efficiency_2016,heidenreich_simulations_2007} is  not needed. 

Simulations are carried out for static homogenous electric field strengths of 0.5, 1, 2 and 10~kV/cm. To mimic the VMI detection, a plane is placed perpendicular to the direction of the static electric field, and the Cartesian coordinates and velocities of the extracted electrons are recorded upon passing the plane. The plane is placed in a sufficient distance (5~mm) from the droplet center to avoid that ions reach the plane during nanoplasma expansion. To achieve simulation times of 10~ns and more, necessary for the electrons (including latecomers) to reach the plane, each trajectory is subdivided into parts. The first 4~ps covering the laser-cluster interaction, initial nanoplasma expansion and the vast majority of three-body electron-ion recombinations (TBR) is carried out with a short time step of $10^{-4}$~fs. After this initial part, the recombined electrons are removed from the simulation by unifying them with their corresponding host ions, reducing the ion charges accordingly. This ``recombination step'' not only reduces the number of electrons to be further propagated but is also a necessary prerequisite to increase the MD time step, since the recombined electrons with tight orbits around their host ions and binding energies on the order of several eV would otherwise require to continue the simulation with the small time step to guarantee energy conservation. In the subsequent parts of the simulation, the MD time step is increased successively up to about 0.1~fs, as the interparticle distances increase. Correlated electronic decay processes, which could transfer energy from excited He atoms and ions to neighboring atoms or to quasi-free electrons on the picosecond-nanosecond timescale~\citep{ovcharenko_autoionization_2020,kelbg_temporal_2020} and which could give rise to additional features in the electron kinetic energy spectrum, are not included in the simulation.

\section{Results}
In usual photoionization experiments of atomic or molecular beams, the ionization probability per laser shot is kept well below one to avoid blurring of electron VMIs and broadening of TOF peaks by space-charge effects. In contrast, in this experiment a single laser pulse creates a large number of electrons and ions when the nanoplasma is formed; consequently, we may expect that the detected electron distributions are dominated by Coulomb repulsion between the electrons. Only electrons directly emitted by outer ionization during the laser pulse produce an extended distribution that is not affected by the space charge. Thus, we may expect to measure two main components in the electron VMIs; a dominant structure at low energies due to the extraction of nanoplasma electrons by the static electric field, and a more energetic one due to direct laser-induced outer ionization.

Indeed, all measured VMIs containing nanoplasma electrons feature a two-component distribution, as shown in Fig.~\ref{fig1}. It is important to note that the brightness and structure of the electron distributions greatly vary from shot to shot. The top panels a)-d) show selected VMIs for large He droplets composed of $2\times10^5$ He atoms on average, whereas the middle panels e)-h) are for small droplets composed of $2\times10^3$ He atoms. The bright inner circular spot mostly has a rather flat intensity profile and a well-defined outer edge for the case of large droplets. The intensity level inside the inner spot is nearly constant irrespective of the spot size, resulting in high total brightness for large spots.

For small droplets, a great shot-to-shot variation of the structure of the bright inner spot and the ratio of inner spot vs. the diffuse outer distribution is observed. This can be seen in the kinetic energy spectra inferred from panels e-h) by an inverse Abel transformation~\citep{garcia_two-dimensional_2004}, see lower panels in Fig.~\ref{fig1}. The peaks at energies $\lesssim 1~$eV result from the central spot, whereas the extended shoulder at $\gtrsim 2~$eV in the two spectra in the middle is due to the scattered electrons around it. As the latter component is distributed over a large area of the detector, it is notoriously difficult to unambiguously discriminate it from noise. The VMI technique is generally better suited for detecting low-energy charged particles than broad, high-energy distributions, in contrast to field-free time-of-flight methods~\citep{kelbg_auger_2019-1,kelbg_temporal_2020} and electrostatic analyzers. Therefore, in this work we refrain from a detailed analysis of the diffuse energetic electron component. For large He droplets, isolated events were observed where the inner spot deviated from a perfect circular shape, showing dents [panels c) and d)] or slightly oval elongations. These distortions indicate that the shapes of large He droplets deviate from the spherical symmetry, which is well-known from x-ray diffraction experiments~\citep{Gomez:2014,langbehn2018three}.

\begin{figure}
\center
\includegraphics[width=1.0\columnwidth]{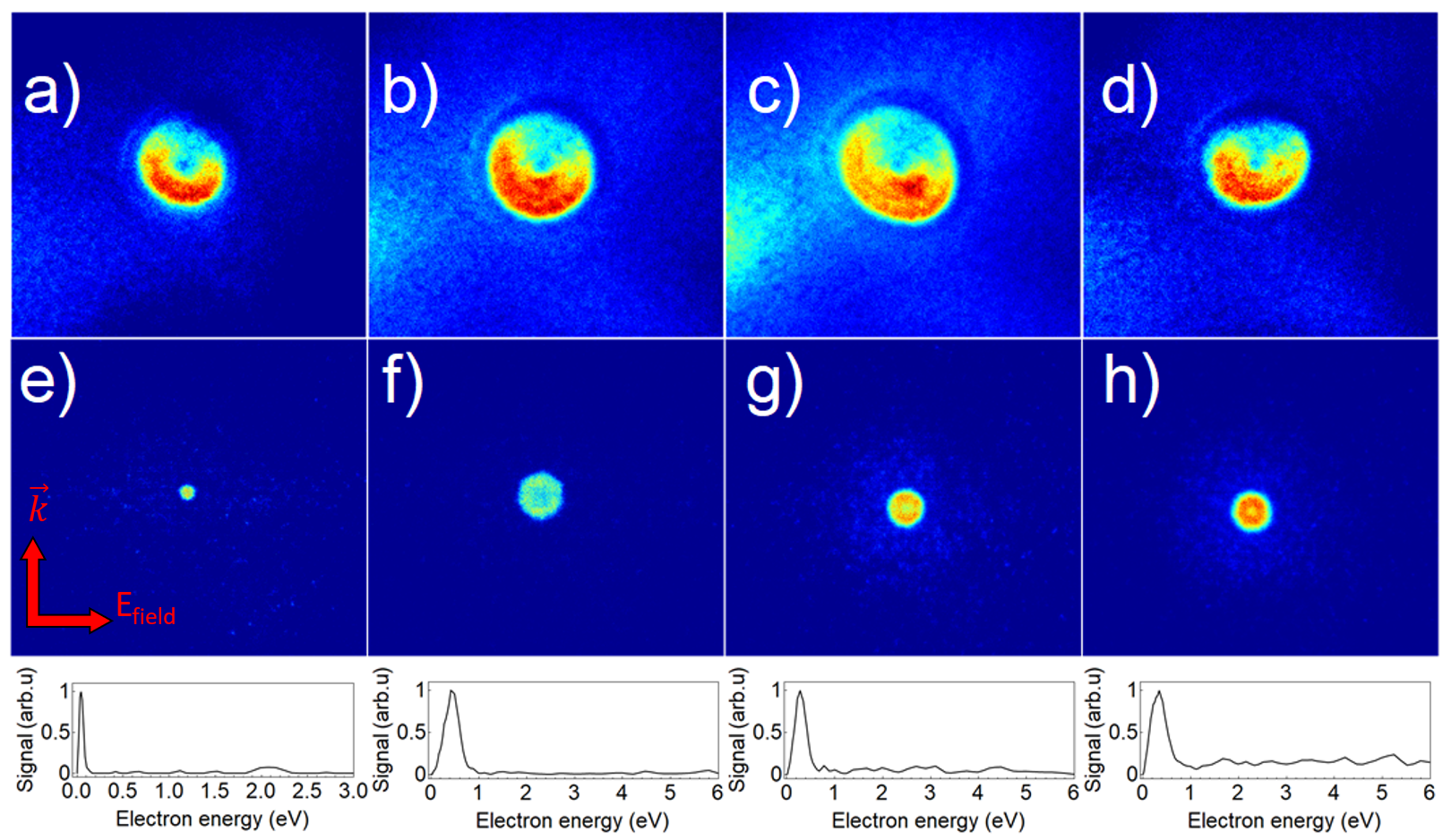} \caption{\label{fig2} Selected single-shot electron images of Ne clusters. a-d) Large Ne clusters consisting of on average $6000$ Ne atoms doped with 10 Xe atoms ionized by a MIR laser pulses. e-h) Ne$_{2000}$Xe$_8$ and their respective electron spectra (bottom row) ionized by NIR pulses.}
\end{figure}

Similar experiments were carried out with Xe-doped Ne clusters, see Fig.~\ref{fig2}. In panels a-d), Ne clusters with an estimated mean size of $N_\mathrm{Ne}=6\times10^3$ atoms were irradiated by intense MIR laser pulses; panels e-h) show VMIs of Ne clusters of size $N_\mathrm{Ne}=2\times10^3$ irradiated by NIR pulses. The general structure of the VMIs is very similar to that of He droplets: a bright inner spot surrounded by a diffuse cloud of more energetic electrons with variable brightness and radius. While small Ne clusters mostly generated circular inner spots, for large Ne clusters the inner feature tends to significantly deviate from the circular shape.  This is likely a direct manifestation of a non-spherical shape of the Ne clusters and should be studied in more detail both experimentally and theoretically. The nonuniform shape of the diffuse cloud around the inner spot in a)-d) is due to technical issues with the spectrometer present in these measurements. Likewise, the low-brightness region in the upper part of the inner spot is due to a reduced sensitivity of the detector. When bright nanoplasma VMIs are recorded at a high repetition rate, degradation of the detector is a serious issue. The electron spectra inferred from the VMIs in e)-h), shown in the bottom row, have a similar structure as the electron spectra of small He nanodroplets. This indicates that single-shot VMIs of nanoplasma electrons are not particularly element specific but rather reflect the generic dynamics of electron emission out of a nanometer-sized quasi-neutral expanding plasma. 

\begin{figure}
\centering
\includegraphics[width=1.0\columnwidth]{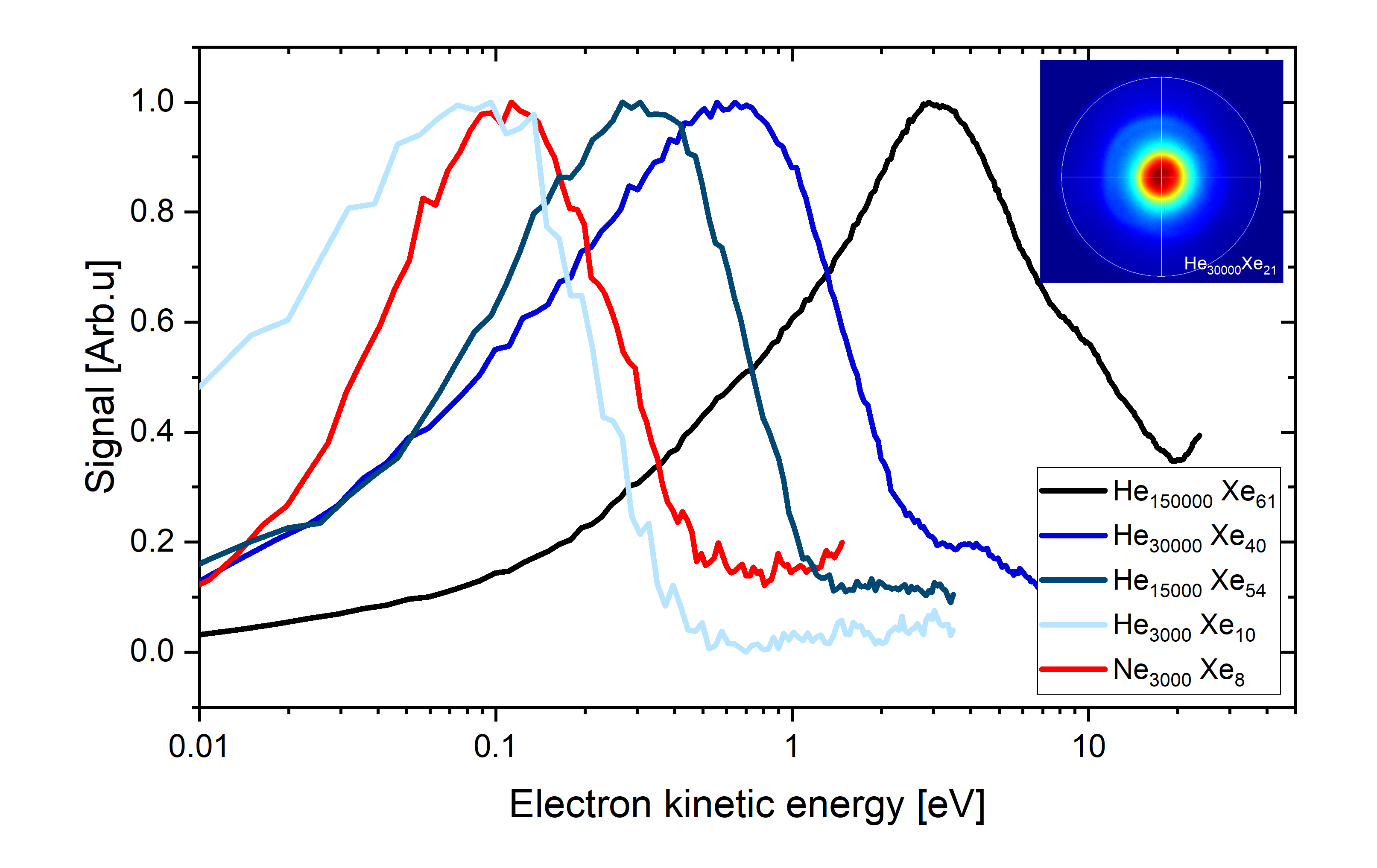} \caption{\label{plothisto1} Electron spectra of averaged VMIs for different He cluster sizes compared to Ne clusters. The inset shows an example of the averaged VMIs for He droplets of size $3\times10^4$ He atoms doped with on average 40 Xe atoms. The outermost white circle indicates the edge of the imaging detector.}
\end{figure}

Assuming that the VMI technique is in principle applicable to imaging of nanoplasma electrons, we can convert the measured radial intensity profile into an electron kinetic energy distribution according to electron-trajectory simulations validated by calibration measurements~\citep{schomas_compact_2017}. Fig.~\ref{plothisto1} shows electron spectra obtained in this way from averaged VMIs recorded for various mean sizes of He droplets and Ne clusters. For every cluster size, around 100 images containing nanoplasma signals were averaged. As an example, the inset displays the resulting average image for He droplets of mean size $N_\mathrm{He}=3\times10^4$ atoms.

The electron spectra show a pronounced peak structure where the peak position shifts from 0.1 up to 3~eV as the mean He droplet size increases from $N_\mathrm{He}=3\times10^3$ $2\times10^5$. The electron spectrum for Ne$_{3000}$ clusters doped with on average 8 Xe atoms closely resembles the one of He$_{3000}$ doped with 10 Xe atoms. Thus, by varying the mean cluster size in the shown range, the kinetic energy of emitted nanoplasma electrons shifts by more than a factor of 10.

More detailed information should be obtained by analyzing the individual single-shot VMIs. This was done in two steps. First, those images containing clear signals from nanoplasma electrons were identified by integrating a small area around the center of the image and comparing the result with a fixed threshold value. In most of the measurements, a maximum of 10\,\% of images contained hits and the rest contained only small numbers of electrons emitted from the residual gas or from small He clusters in the size range 100-1000 atoms. Second, the radius of the inner bright spot was determined by a simple algorithm that computes the derivative of the radial intensity profile. The radius is then defined by the largest negative value of the radial derivative. This radius value is directly converted into energy corresponding to the maximum of the distribution, $E_\mathrm{max}$. The integral of the area within the radius is converted into the total number of electrons by analyzing the brightness of single electron hits on the detector measured in the background images.


\begin{figure}
\centering
\includegraphics[width=1.0\columnwidth]{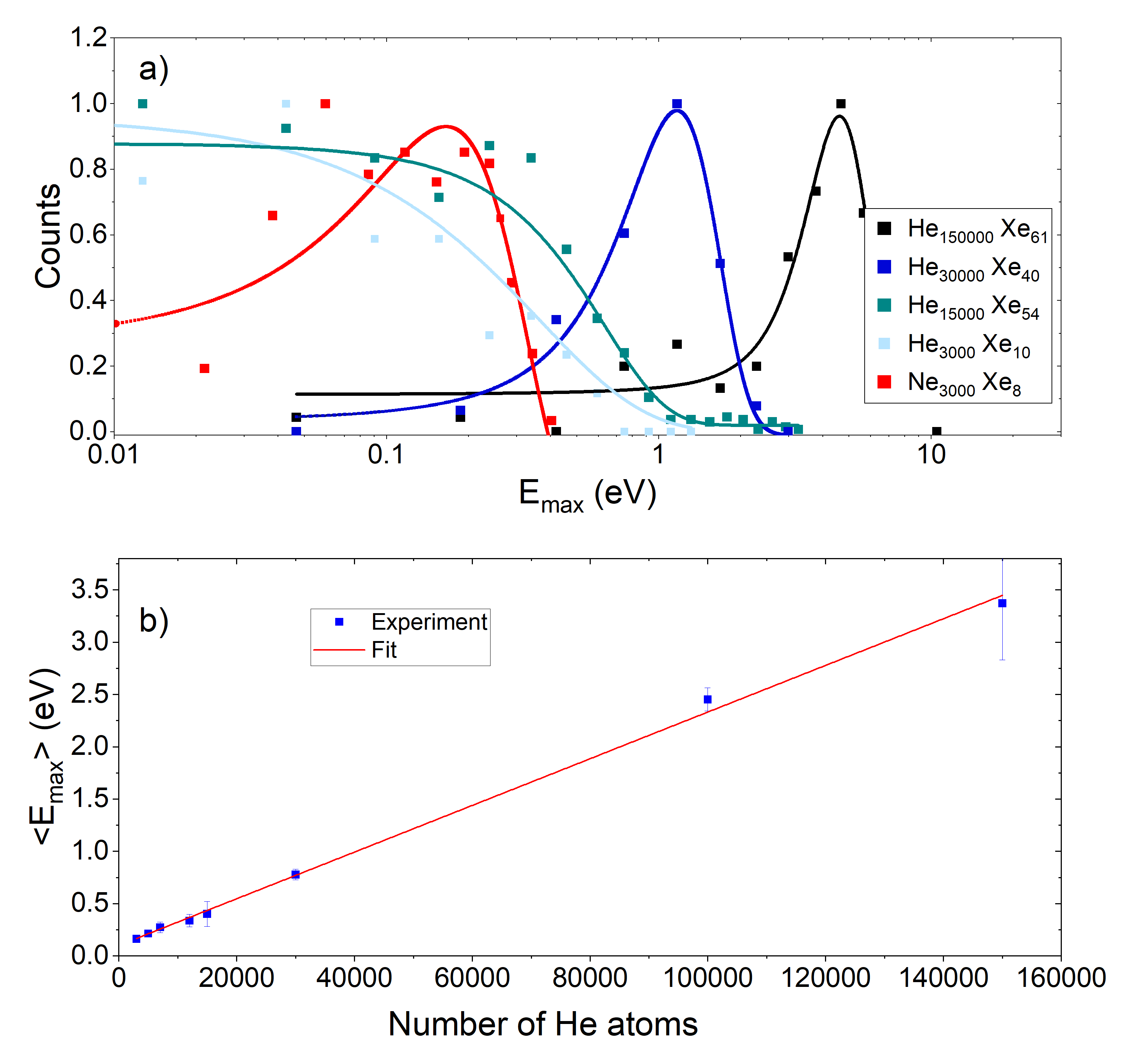}  \caption{\label{plothisto2} a) Histograms of the electron peak energies $E_\mathrm{max}$ of the single-shot VMIs that were averaged in Fig.~\ref{plothisto1}. The data points are the energies for the different He and Ne clusters sizes. The solid lines are Gaussian fits. b) Cluster size dependence of the mean peak energy for the He data found in a) including additional cluster sizes. The red line is a linear fit of the experimental data.}
\end{figure}

Fig.~\ref{plothisto2} a) shows the resulting histograms of $E_\mathrm{max}$ values for the individual shots used in Fig.~\ref{plothisto1}. The solid lines are fits by a Gaussian function. Overall, both analysis methods are in good agreement. For large He droplets, the peaks in Fig.~\ref{plothisto2} a) are slightly narrower than those in Fig.~\ref{plothisto1}. This is expected, as in the histogram analysis [Fig.~\ref{plothisto2} a)], the peak width is solely determined by the scatter of $E_\mathrm{max}$, whereas the shape of each individual electron spectrum is not accounted for. For small clusters, the peaks extend further towards low energies $E_\mathrm{max}<0.1~$eV, indicating that a large number of VMIs with small inner spots are present, which contribute less to the averaged images (Fig.~\ref{plothisto1}). The mean values of the $E_\mathrm{max}$ distributions shown in a) and from additional measurements are depicted as squares in Fig.~\ref{plothisto2} b). As shown by the solid line, these mean values $\langle E_\mathrm{max}\rangle$ closely follow a linear dependence of the number of He atoms. 

\begin{figure}
\center
\includegraphics[width=1.0\columnwidth]{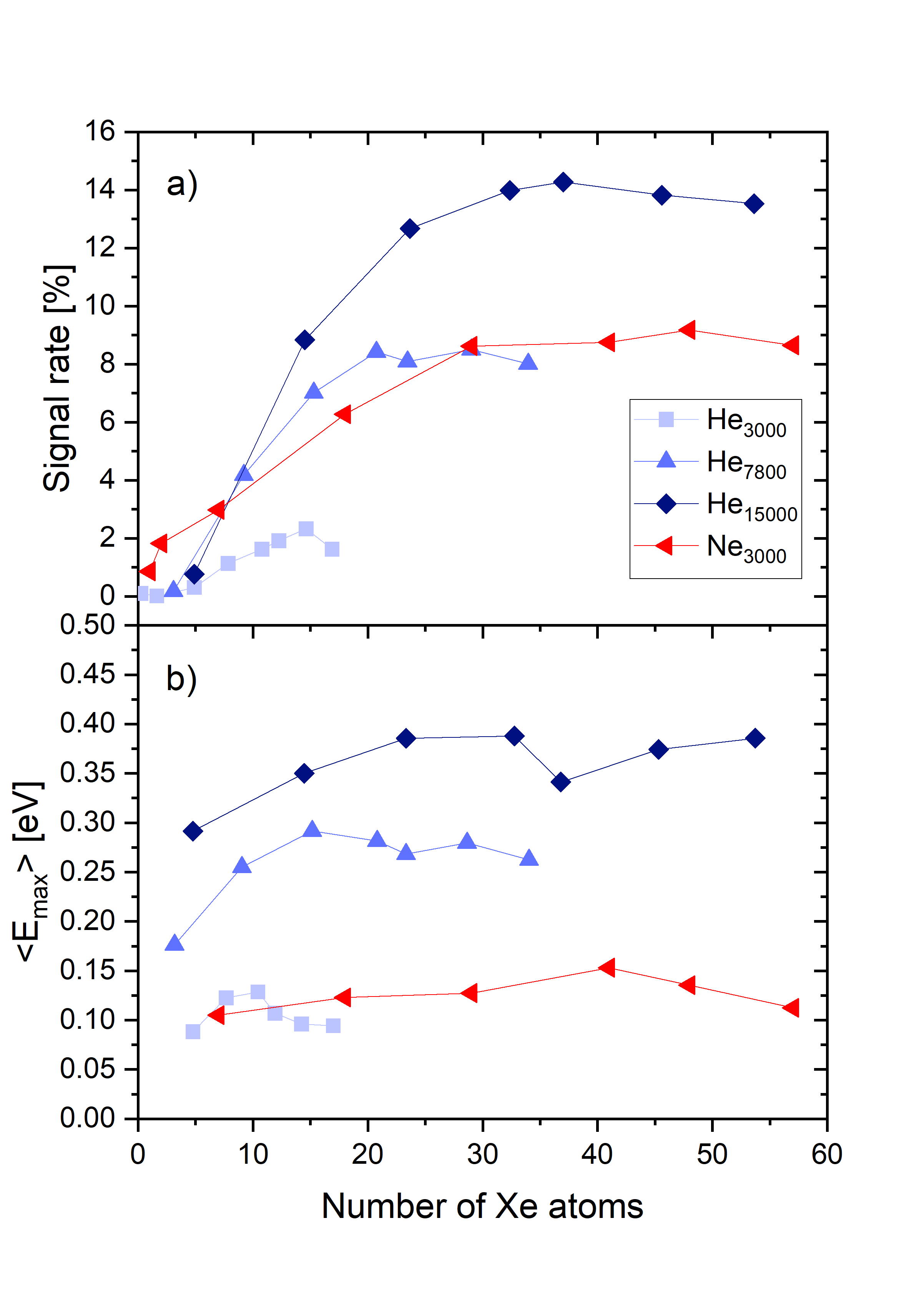}  \caption{\label{plotDop} Dependence on the number of dopant atoms. a) Fraction of VMIs containing nanoplasma electrons for different He and Ne cluster sizes ionized by NIR pulse. b) Mean electron peak energy. The blue symbols correspond to He droplets of different sizes, the red symbols correspond to Ne$_{3000}$.}
\end{figure}

When we apply the same procedure to VMIs recorded for varying Xe partial pressure in the doping cell, we obtain the data shown in Fig.~\ref{plotDop} b) for He and Ne clusters of different sizes. The data show no significant variation of the nanoplasma electron energy as a function of the number of Xe dopant atoms embedded into the clusters. Thus, the nanoplasma is primarily determined by the size of the He or Ne host cluster, whereas the Xe dopant cluster only facilitates the ignition of the nanoplasma, as previously shown~\citep{mikaberidze_laser-driven_2009,krishnan_dopant-induced_2011,Heidenreich:2016,heidenreich_dopant-induced_2017}. For higher doping levels, scattering and destruction of the He droplets by multiple collisions with Xe atoms in the doping cell reduces the signal rate~\citep{Heidenreich:2016}. A minimum of about 5 Xe atoms is needed to ignite a He nanoplasma, whereas Ne nanoplasmas are measured even at the lowest doping levels of 1 Xe dopant atom per cluster on average. This is due to the lower ionization energy of Ne, which makes pure Ne clusters more prone to strong-field ionization as compared to He nanodroplets. The overall similar behavior of Ne clusters and He nanodroplets with respect to their doping dependence of the nanoplasma ignition indicates that the structure of the cluster (solid vs. superfluid) is not decisive for the dopant-induced activation of a nanoplasma.

\begin{figure}
\center
\includegraphics[width=1\columnwidth]{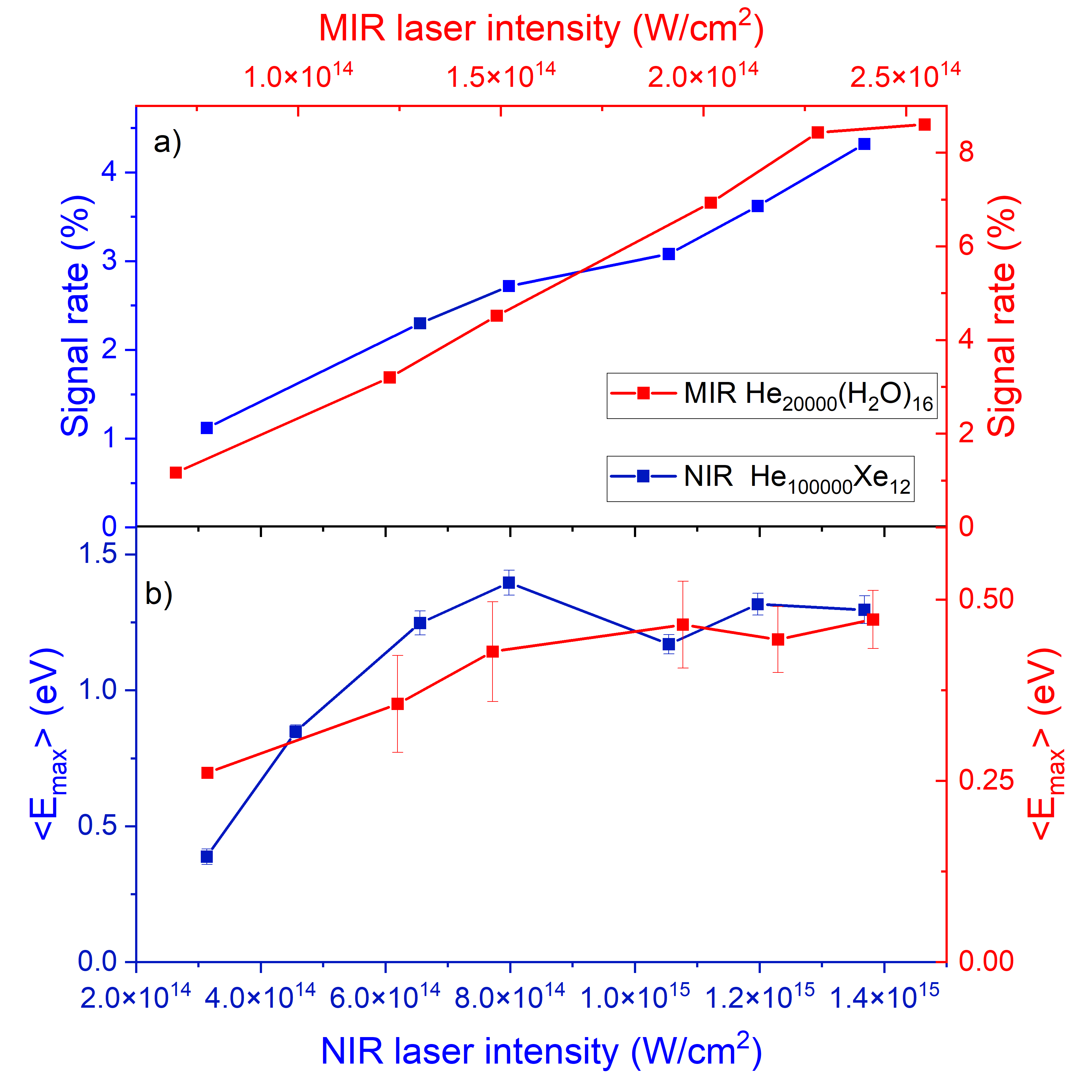} \caption{\label{Intens} Laser intensity dependence. a) Nanoplasma hit rate for He droplets irradiated by NIR and MIR pulses at different intensities. b) Electron peak energy for the two laser types. Notice that the blue axis (bottom and left) corresponds to the NIR data and the red (top and right) to the MIR data.}
\end{figure}

Similar to the doping dependence, the intensity of the laser pulses significantly influences the abundance of nanoplasma events, whereas the mean electron energy is only weakly affected. Fig.~\ref{Intens} shows the intensity dependence for He nanodroplets doped with Xe atoms exposed to NIR laser pulses and for He droplets, doped with water, exposed to MIR pulses at different intensities (note the colored axes on the top and bottom of the figure). In both cases, the signal rates follow a nearly linear rise with laser intensity. However, the mean electron energy remains nearly constant within the experimental uncertainty. This is due to intensity averaging over the volume of the laser focus; while a few droplets are hit by the laser in the center of the focus and generate more energetic electrons at higher laser intensity, a large number of droplets in the periphery of the focal volume experience a reduced local intensity, which may still be sufficient for igniting a nanoplasma. These droplets lead to an enhanced signal rate with rising laser intensity but contribute mostly with low energies. Owing to the single-shot method, we can select only the brightest nanoplasma images thereby selecting events where the largest droplets are exposed to the peak intensity. These selected images (not shown) clearly display enhanced electron numbers and energies when the intensity increases.

\begin{figure}
\center
\includegraphics[width=1\columnwidth]{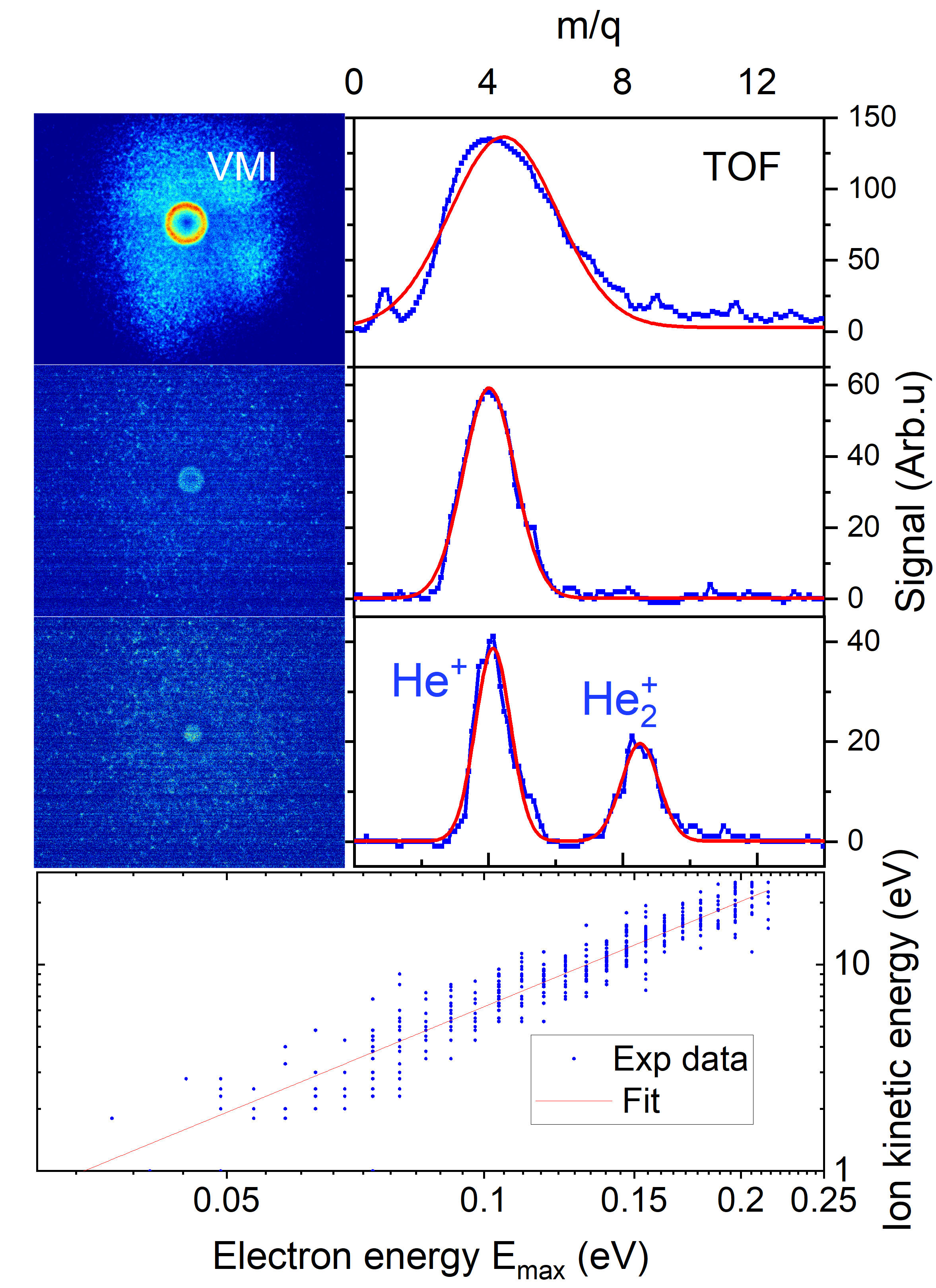} \caption{\label{vmi-tof} Selected single-shot electron VMIs of He droplets of size $10^4$ He atoms doped with calcium atoms, irradiated by MIR pulses and ion TOF mass-over-charge spectra measured simultaneously. Blue lines are the experimental data, red lines are Gaussian fits. The bottom panel represents ion kinetic energies inferred from the fits vs. electron peak energies for individual correlated electron VMIs and ion TOF traces. The red line is a power-law fit. }
\end{figure}
One advantage of the electron VMI technique is the possibility to combine it with ion detection. To demonstrate this capability, we ignited He nanoplasmas from calcium-doped He nanodroplets
with intense MIR pulses and measured electron VMIs and ion TOF traces for every hit. The upper part of Fig.~\ref{vmi-tof} shows typical electron VMIs and correlated ion TOF traces. The bright feature in the VMIs (large number of emitted electrons) clearly correlates with an intense and broadened He$^+$ peak indicative of high ion kinetic energies. Even fully ionized He atoms, He$^{2+}$, are visible as a small peak in the TOF spectrum shown in the top right panel. The fact that it appears at a value of $m/q$ slightly below 2 is due to the high He$^{2+}$ kinetic energy causing peak deformations~\citep{krishnan_dopant-induced_2011,krishnan_evolution_2012,Heidenreich:2016,heidenreich_dopant-induced_2017}. In contrast, low-intensity VMIs correlate with narrow He$^{+}$ and He$^{+}_2$ peaks in the TOF traces, indicating low ion kinetic energies and incomplete fragmentation or ion dimerization occurring in a nanodroplet that is only partly ionized (bottom VMIs panels). The red lines are Gaussian fits to the peaks. 

The widths $\sigma$ of the He$^{+}$ peak fits are converted into ion kinetic energies $E_i$ by the following procedure: Using the SIMION simulation software, TOF traces were simulated for a large number of ion kinetic energies $E_i'$ in the range 0-30~eV following an exponential distribution $\propto\exp(-E_i'/\tau)$ [cf.~Fig.~\ref{vmi-theo} d)]. At higher ion energies $E_i'>25$~eV, the efficiency of our detector significantly drops because energetic ions tend to miss the sensitive area of the detector. Therefore, at $E_i\gtrsim10$~eV the reported ion energies have to be considered as lower bounds. The simulated He$^+$ TOF distributions can be fitted to the measured peaks, thereby providing an empirical relation between the TOF peak widths $\sigma$ and $\tau\equiv E_i$. 
The bottom panel of Fig.~\ref{vmi-tof} shows $E_i$ for He$^+$ vs. electron peak energies inferred from the corresponding electron VMIs. Each blue dot represents one measurement of a correlated electron VMI and ion TOF trace. The red line is a fit of the power law $E_\mathrm{max}\propto E_i^x$ resulting in $x=1.65$. This model predicts $E_i=290$~eV for $E_\mathrm{max}=1$~eV and $E_i\sim10^4$~eV for $E_\mathrm{max}=10$~eV.

\section*{Discussion}
The evolution of a laser-induced nanoplasma is a complex dynamical phenomenon that involves various processes of light-matter interactions and collisions on the length scale of atoms and of the cluster as a whole. These include tunnel ionization, inverse bremsstrahlung, electron-impact ionization, plasmon-enhanced resonant absorption, three-body recombination, etc.~\citep{saalmann_mechanisms_2006,fennel_laser-driven_2010}. Accordingly, sophisticated numerical model calculations, such as MD simulations, are usually required to capture the most important aspects of the nanoplasma dynamics.

Nevertheless, we attempted to come up with a simple analytic model to reproduce the most pronounced correlation visible in the VMIs despite their strong shot-to-shot variation, namely that of size and brightness of the inner spot. This model is based on the expansion of a homogeneous spherical distribution of charges of radius $R$ (electrons in our case) driven by Coulomb repulsion. According to Islam~\textit{et al.}~\cite{islam_kinetic_2006}, the initial radial charge density distribution
\begin{equation}
\frac{dP}{dr}=\frac{3 r^2}{R^3} \Theta (R-r)
\label{density_distribution}
\end{equation}
transforms into a final kinetic energy distribution
\begin{align}
\frac{dP}{dE}=\frac{3}{2} \sqrt{\frac{1}{E_R}} \frac{1}{E_R}\sqrt{E} \cdot \Theta\left( 1-\frac{E}{E_R}\right),
\end{align}
where 
\begin{equation}
   E_{R}=\frac{e^2}{4 \pi \epsilon_0} \left(\frac{4}{3} \pi \rho\right)^{1/3} N^{2/3}
   \label{model}
\end{equation}
is the cut-off energy. Here, $\Theta$ is the Heaviside step function, $N$ is the number of electrons in the sphere and $e$ is the elementary charge. In this way, we obtain a simple power law dependence $E_R\propto N^{2/3}$ under the assumption of a constant initial density $\rho=3N/(4\pi R^3)$. Here, $E_R$ can be identified with $E_\mathrm{max}$ in the experimental electron spectra and $N$ is derived from the integrated brightness of the inner spot. 
\begin{figure}
\center
\includegraphics[width=1.0\columnwidth]{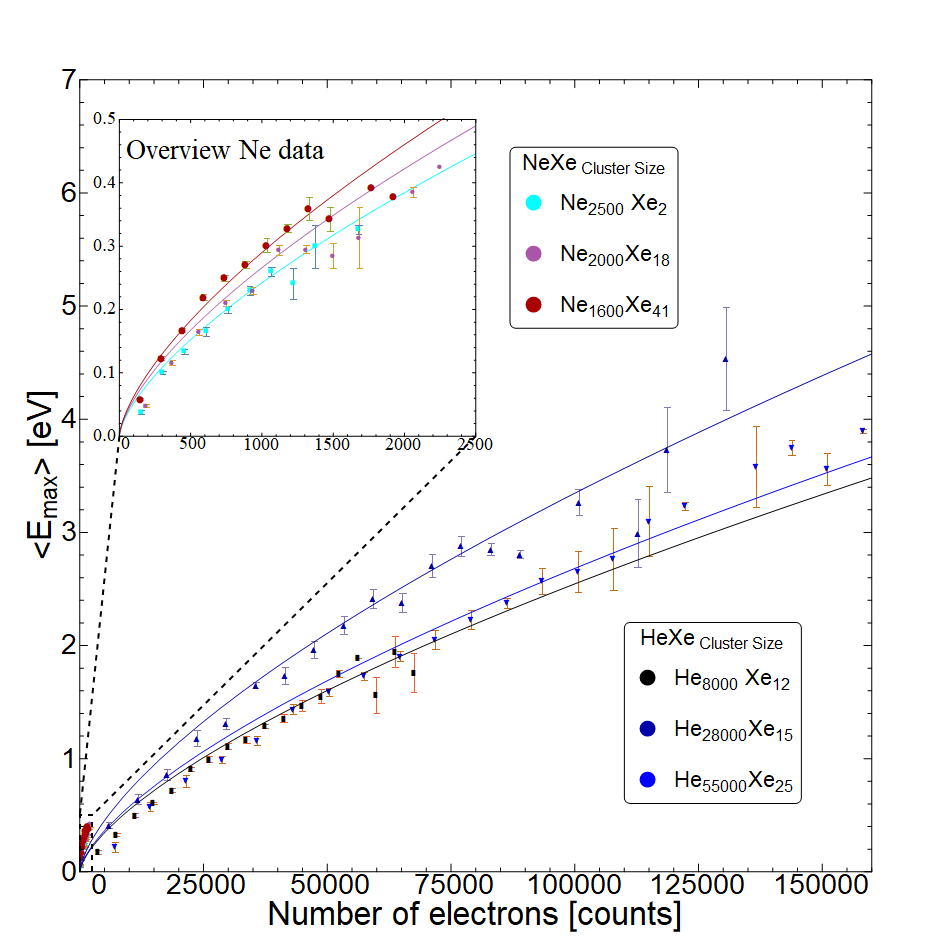}\caption{\label{plotfit} Peak electron energies as a function of the number of nanoplasma electrons. The data points represent the mean values of the distributions of electron peak energies for different He and Ne cluster sizes. The solid lines represent the fits based on Eq.~\ref{model}. The inset shows a closeup of the low-energy region where the Ne data points are concentrated.}
\end{figure}

The correlation of $E_\mathrm{max}$ and $N$ is represented in Fig.~\ref{plotfit}. The data points are obtained by binning $E_\mathrm{max}$ and $N$ values from a total of about $1000$ VMIs for every He and Ne cluster size. The error bars indicate the dispersion of the data within each bin, resulting from 10-200 VMI measurements per data point. The solid lines are fits of Eq.~\ref{model} to the data. Surprisingly, despite of the simplicity of our model, the experimental data for both He and Ne clusters are well reproduced by the fits. Due to the limited range of sizes of Ne clusters accessible in the experiment, the Ne data were constrained to $N\lesssim 3\times10^3$. The inset shows an overview of the data and fits at low energies.

The only free fit parameter is the prefactor in Eq.~\ref{model} which yields $\rho= 0.05$-$0.2\,e\,\mu$m$^{-3}$ for He nanodroplets, where $e$ is the electron charge. The smallest value is obtained for small droplets at low NIR intensity $I\sim 5\times10^{14}~$Wcm$^{-2}$, the largest value is obtained for large droplets and $I\gtrsim 10^{15}~$Wcm$^{-2}$. These values are much smaller than the atom density of He nanodroplets, 0.022~\AA$^{-3}$. This is expected in view of our tentative interpretation that the relevant electron density responsible for the lower-energy component is the one in the more dilute electron beam extracted by the external static electric field of the VMI spectrometer, see below. In the hotter nanoplasma created by more intense pulses in large droplets, the electrons appear to be extracted more easily, \textit{i.~e.} at an earlier stage of the nanoplasma expansion when the electron density is higher. While the insight into the physics of nanoplasmas gained from this model certainly is rather limited, it is still useful for predicting the energy of nanoplasma electrons merely based  on the detection of their yield. 

\begin{figure}
\center
\includegraphics[width=1\columnwidth]{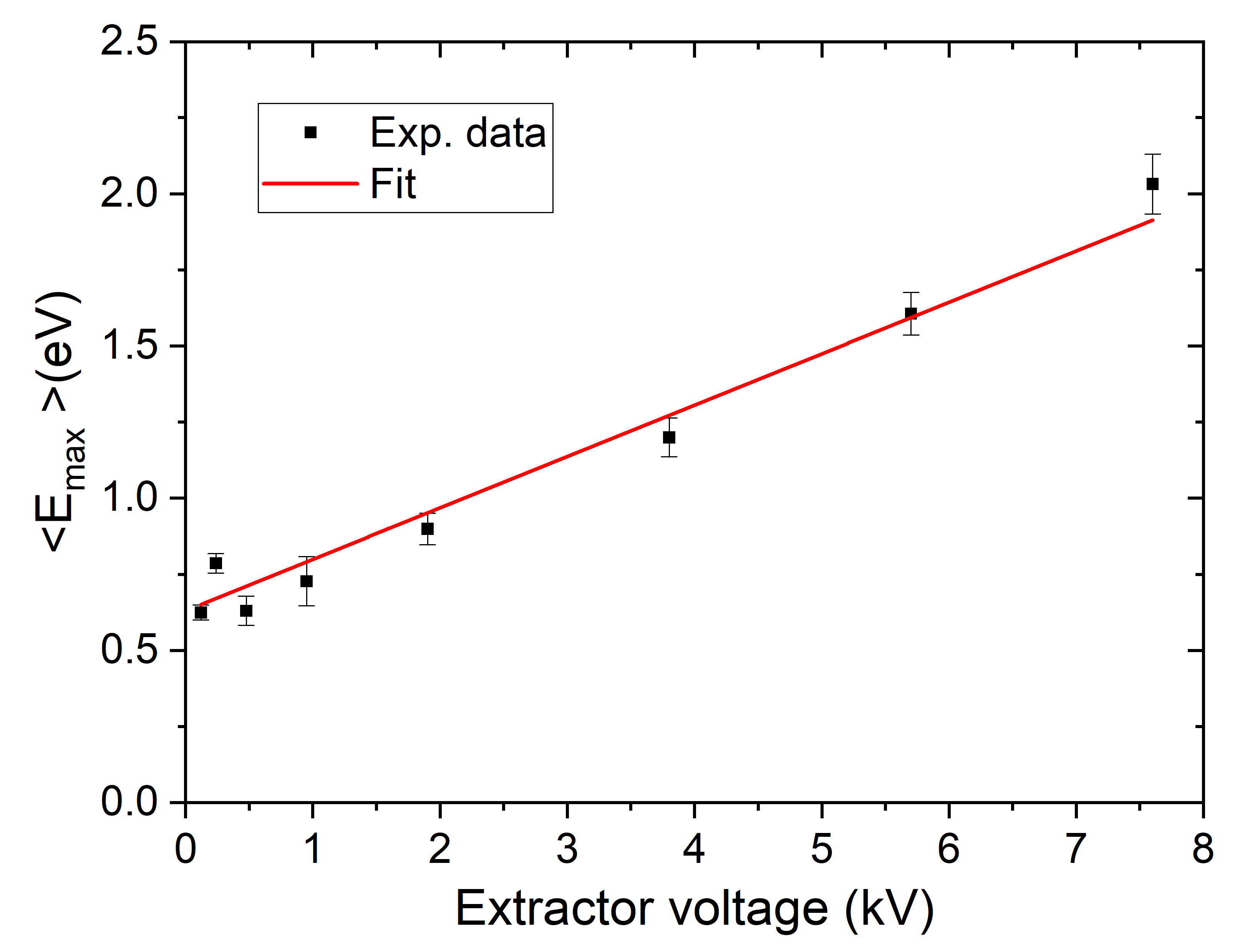} \caption{\label{vmi-volt}Dependence of the mean electron peak energy on the extractor voltage for He$_{8000}$Xe$_{12}$ irradiated by NIR pulses. The red line is a linear fit of the data.  }
\end{figure}

Up to this point we have discussed our results assuming that the VMI technique applies when detecting electrons emitted from a nanoplasma. However, we find indications that the evolution of a nanoplasma, in particular the emission of electrons, is affected by the static electric field present in the VMI spectrometer. 
Fig.~\ref{vmi-volt} shows the mean electron peak energies as a function of the voltage applied to the extractor electrode of our spectrometer. The electric field (in V/cm) in the interaction region is given by a factor $0.2$ times the extractor voltage (in V). We see a clear trend of increasing detected electron energies for higher electric fields, roughly following a linear dependence. The extrapolation toward zero electric field indicates a finite value of the electron energy which is a factor of 0.7 lower than the values presented above, measured at an extractor voltage of 1.9~kV. Thus, to compare with field-free measurements or simulations, the reported energies should be scaled accordingly.

\begin{figure}
\center
\includegraphics[width=1\columnwidth]{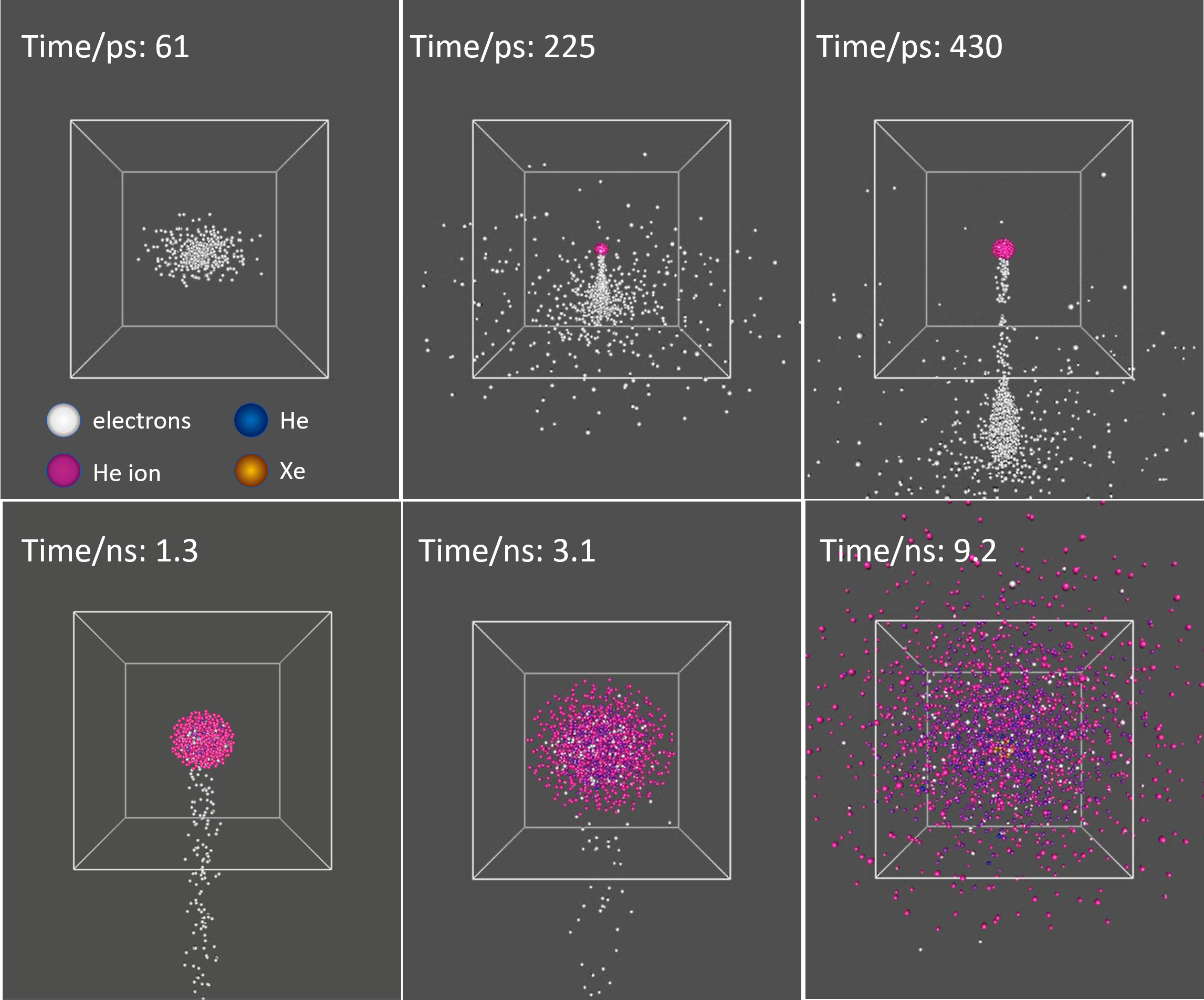} \caption{\label{theo} Theoretical time-evolution of a nanoplasma ignited by an intense NIR laser pulse in the presence of an external homogeneous electric field ($F = 1000$~V/cm). The laser polarization is horizontal and the electric field points upwards.}
\end{figure}

To get a better understanding of the impact of a static electric field on the evolution of a nanoplasma and on the extraction of electrons out of it, we have carried out MD simulations with and without an external homogeneous electric field. Fig.~\ref{theo} shows snapshots of the simulation at various times after the maximum of the laser pulse. A NIR laser pulse of 35~fs duration (FWHM) and a maximum intensity of $2\times10^{14}~$Wcm$^{-2}$ was applied to a He nanodroplet consisting of 2171~He atoms doped with 23 Xe atoms. The polarization of the laser is horizontal, and the electric field $F=500$~V/cm points upwards. Electrons, neutral He and Xe atoms as well as ions are represented by differently colored beads. 

In the early stage $t\lesssim 430$~ps (top frames), electrons (white dots) are emitted in two rather distinct distributions; a diffuse and nearly isotropic cloud is created shortly after the laser pulse (see frame at 61~ps), and one highly directional component of electrons is extracted downwards along the electric field axis at $t\gtrsim 100$~ps. At $t>1~$ns (bottom frames), the expanding cloud of ion cores (pink beads) slowly move upwards along the electric field.
\begin{figure}
\center
\includegraphics[width=1\columnwidth]{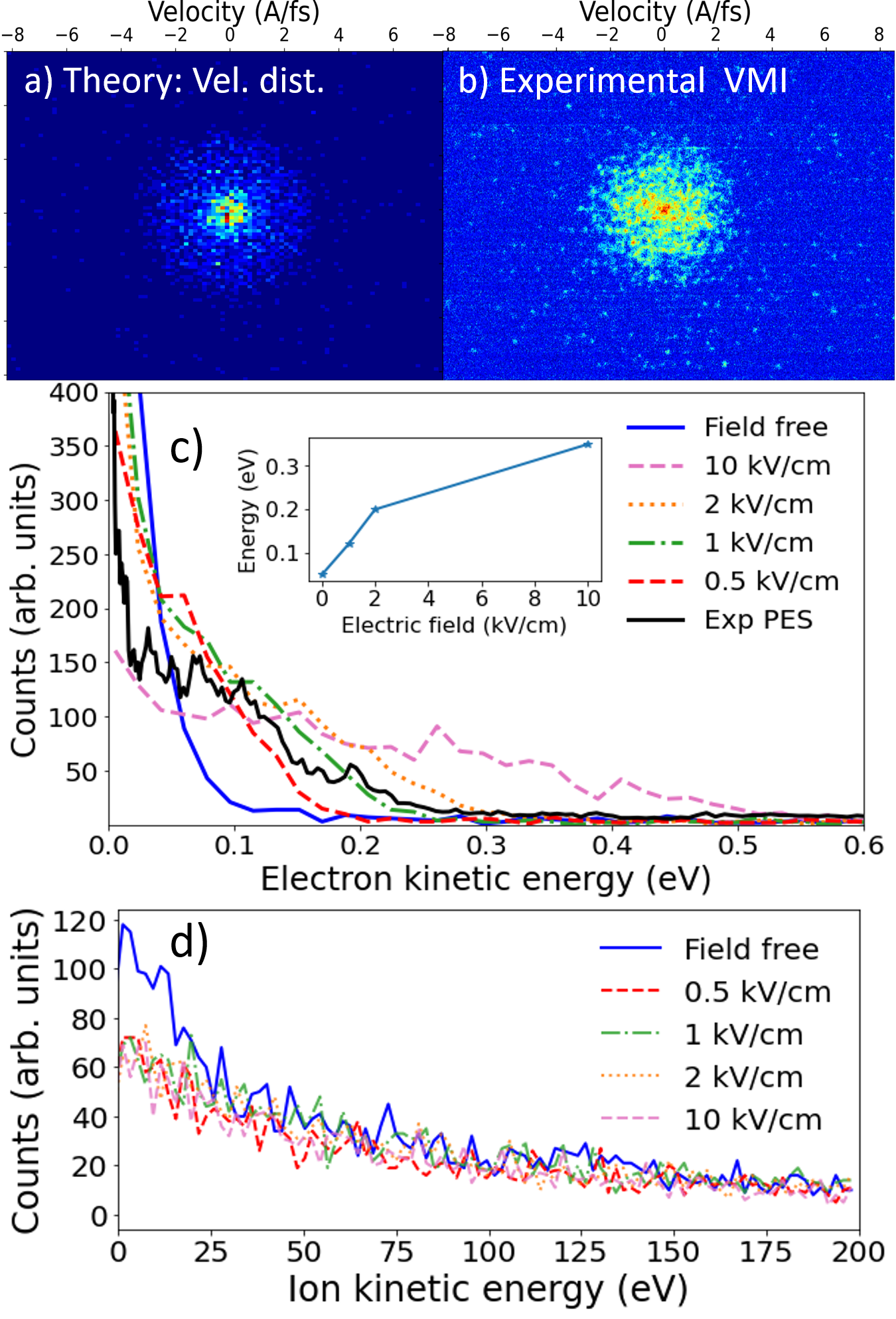} \caption{\label{vmi-theo}
Comparison of theoretical and experimental results. a) Simulated electron velocity distribution for a doped droplet, He$_{2171}$Xe$_{23}$,  exposed to a NIR pulse in an electric field of $F=1000$~V/cm. b) Selected experimental electron VMIs for He$_{8000}$Xe$_{20}$ ionized by NIR pulses in the presence of an electric field of 25~V/cm in the interaction region. c) Electron kinetic energy distributions from the simulation (colored lines) and the experiment (black line). d) Corresponding simulated ion kinetic energy distributions.}
\end{figure}

From these simulations we can extract both spatial and velocity coordinates of electrons as they pass through a plane located 5~mm below the droplet. Since the electric field in the simulation is homogeneous, the VMI condition is not fulfilled. Nevertheless, we can compare the distribution of velocity components perpendicular to the electric field with the measured VMIs, see Fig.~\ref{vmi-theo}. The VMI shown in panel b) is a typical instance of a class of images where the bright central spot has a diffuse edge [cf. Fig.~\ref{fig1} f)]. Such images make up about 5~\% of all hits. Indeed, the simulated distribution of transverse velocities, shown in Fig.~\ref{vmi-theo} a), resembles the experimental VMI in many respects. It shows two main components: A speckled background of fast electrons, and a more compact circular cloud in the center. In addition, a small bright dot in the center indicates a small contribution of electrons with nearly vanishing energy. Note that the correspondence of simulated and measured distributions holds even in absolute terms, see the velocity scales on the top of panels a) and b). We take this as a confirmation of our interpretation of the central spot as resulting from the extraction of nanoplasma electrons by the external electric field. 
Fig.~\ref{vmi-theo} c) shows the distributions of the simulated transverse electron kinetic energies for electric field strengths $F=0$, 0.5, 1, 2, and 10~kV/cm as colored lines. Clearly, an increasing electric field gives rise to a growing shoulder structure extending up to 0.6~eV at $F=10$~kV/cm (blue dashed line), which is not present without electric field (red line). Electron energies inferred from this shoulder are depicted in the inset. The black line depicts the electron spectrum obtained from the Abel inversion of the average of 5 images of the same type as that shown in panel b), arbitrarily scaled in intensity. It qualitatively resembles the simulated spectra at low electric field. However, the electron energies inferred from the simulated spectra fall significantly below the average of experimentally determined electron energies (Fig.~\ref{vmi-volt}), despite the higher electric fields used in the simulation. This deviation is likely related to the approximations made in the MD simulations, cf. Sec.~\ref{sec:MD}.

Fig. 12 d) shows the ion kinetic energy distributions derived from the simulated ion velocity components perpendicular to the electric field, for several values of $F$ and for the field-free case. The energy distribution of the field-free simulation (blue curve) is distinguished by a significantly stronger low-energy component in the range 0-20 eV. A low-energy component in the ion kinetic energy usually indicates the presence of an electron-rich nanoplasma core~\cite{Last:2006,Heidenreich:2012}. The partial suppression of the low-energy component in the ion energy spectra is consistent with the extraction of nanoplasma electrons by the static electric field. 

Clearly, this study should be further extended to fully characterize the electric-field dependence of electron emission from a nanoplasma in various regimes of laser intensity and droplet size. The fact that the simulated electron spectra do not reproduce the exact peak structure in most of the experimental spectra at energies exceeding 1~eV (Fig.~\ref{fig1} and \ref{fig2}) is likely due to the relatively small size of the He droplet used in the simulation (He$_{2171}$Xe$_{23}$). Larger droplets emit larger amounts of electrons at higher densities which likely give rise to a more pronounced shoulder in the spectra, eventually evolving into a peak. Unfortunately, simulations of large He droplets are prohibitively costly for this work. Additionally, the inhomogeneous electric field of the VMI spectrometer may cause distortions of the velocity distributions due to the coupling of longitudinal and transverse degrees of freedom with respect to the electric field. 

\section{Conclusion}
In conclusion, we presented the first dedicated experimental study of single-shot VMI of electrons emitted from He and Ne nanoplasmas induced by intense NIR and MIR laser pulses. The images are characterized by large shot-to-shot fluctuations of the brightness and structure due to the large variation of cluster sizes and of laser intensities seen by each cluster as it is hit at different positions within the laser focus. This puts some demands on the data storage capacities, sorting and analysis methods to cope with the large amount of image data. To clearly map systematic dependencies, typically a minimum of about 100 images have to be recorded per set of parameters. Despite the rather high intensities used to strong-field ionize the clusters, the electron emission patterns are circularly symmetric to a high degree. Only slight dents and elongations along the laser polarization were observed for the highest intensities $\gtrsim 10^{15}~$Wcm$^{-2}$. The VMIs of both He and Ne clusters feature a generic two-component structure consisting of a central bright spot and a diffuse cloud of electrons around it. Likewise, VMIs measured for MIR and NIR laser pulses are very similar, which supports the concept that electron emission is mostly determined by the intrinsic dynamics of the nanoplasma. The initial structure of the clusters (superfluid, solid) and the characteristics of the laser pulses only play minor roles once the threshold for avalanche ionization is reached. The most sensitive parameter that determines the yield and energy of emitted electrons is the cluster size.

We have demonstrated the possibility to record electron VMIs and ion TOF traces simultaneously on a shot-to-shot basis. Bright electron images clearly correlate with large yields of singly and doubly charged He atomic ions, whereas low nanoplasma electron yields correlate with singly charged He$^+$ and He$_2^+$ ions. Electron and ion kinetic energy are found to be related by a simple power law. Despite the large shot-to-shot variations, a clear correlation of the brightness (number of nanoplasma electrons) and size of the inner spot (maximum electron energy) is found. It is consistent with a simple power law derived for an expanding spherical cloud of electrons. A more realistic MD simulation including a static electric field unravels the mechanism of electron extraction out of the expanding nanoplasma. It shows that the resulting transverse kinetic energy distributions of electrons significantly differ from the electron spectrum obtained in the case of field-free expansion in that they feature a shoulder structure around $0.5$~eV. Thus, simulated electron spectra reproduced the experimental spectra well for a certain class of events. The results of the simulations are supported by VMI measurements at varying spectrometer voltages. The transverse electron energy inferred from the electron images clearly depend on the spectrometer electric field, consistent with the result of the simulation. 

In general, the VMI technique has the advantage of collecting all electrons emitted into the full solid-angle, which makes it sensitive enough that entire electron spectra can be measured for individual clusters and nanodroplets. Moreover, with this technique, electrons and ions can be detected simultaneously giving more insight into the ionization dynamics. However, the concept of extracting accurate kinetic energies by inverse Abel transformation of the measured images breaks down because the spectrometer field interferes with the expansion dynamics of the nanoplasma. To what extent the measured electron distributions can be reliably related to the field-free electron spectra or to other characteristics of the nanoplasma (degree of inner ionization, electron temperature, initial cluster shape, etc.) remains to be further investigated both experimentally and by means of simulations. Likewise, the effect of the spectrometer field on the nanoplasma dynamics should be elucidated for different regimes of nanoplasma formation such as extreme-ultraviolet and x-ray cluster ionization~\citep{ovcharenko_novel_2014,schutte2014rare,schutte2015recombination,Kumagai:2021}.

    The authors are grateful for financial support from the Deutsche Forschungsgemeinschaft (DFG) within the project MU 2347/12-1 and STI 125/22-2 in the frame of the Priority Programme 1840 ‘Quantum Dynamics in Tailored Intense Fields’, from the Carlsberg Foundation and the SPARC Programme, MHRD, India. The ELI-ALPS project (GINOP-2.3.6-15-2015-00001) is supported by the European Union and co-financed by the European Regional Development Fund. A. H. is grateful for financial support from the Basque Government (project ref. no. IT1254-19) and from the Spanish Ministerio de Economia y Competividad (ref. no. CTQ2015-67660-P). Computational and manpower support provided by IZO-SGI SG Iker of UPV/EHU and European funding (EDRF and ESF) is gratefully acknowledged. We would like to thank the MPIK and ELI-Alps teams for their assistance and support during experiments.

\bibliography{MarcelsBib,NIRpaperbib}

\end{document}